\documentclass[sigconf]{acmart}
\usepackage{enumitem}
\usepackage{amsmath}
\pdfoutput=1
\renewcommand\footnotetextcopyrightpermission[1]{} % removes footnote with conference information in first column

\def\BibTeX{{\rm B\kern-.05em{\sc i\kern-.025em b}\kern-.08emT\kern-.1667em\lower.7ex\hbox{E}\kern-.125emX}}
    
\begin{document}

%
% The "title" command has an optional parameter, allowing the author to define a "short title" to be used in page headers.
\title[DHEN: A Deep and Hierarchical Ensemble Network]{DHEN: A Deep and Hierarchical Ensemble Network for Large-Scale Click-Through Rate Prediction}

%
% The "author" command and its associated commands are used to define the authors and their affiliations.
% Of note is the shared affiliation of the first two authors, and the "authornote" and "authornotemark" commands
% used to denote shared contribution to the research.

\author{Buyun Zhang*, Liang Luo*, Xi Liu*, Jay Li, Zeliang Chen, Weilin Zhang, Xiaohan Wei, Yuchen Hao, Michael Tsang, Wenjun Wang, Yang Liu, Huayu Li, Yasmine Badr, Jongsoo Park, Jiyan Yang, Dheevatsa Mudigere, Ellie Wen}
\thanks{*Three authors contributed equally to this research.}
\email{{buyunz, liangluo, xliu1}@fb.com}
\affiliation{%
  \institution{Meta Platforms, Inc. 1 Hacker Way, Menlo Park, CA}
%   \streetaddress{1 Hacker Way}
%   \city{Menlo Park}
%   \state{CA}
%   \postcode{94065}
}
%
% By default, the full list of authors will be used in the page headers. Often, this list is too long, and will overlap
% other information printed in the page headers. This command allows the author to define a more concise list
% of authors' names for this purpose.
\renewcommand{\shortauthors}{Buyun, Liang and Xi, et al.}

%
% The abstract is a short summary of the work to be presented in the article.

\begin{abstract}
Learning feature interactions is important to the model performance of online advertising services. As a result, extensive efforts have been devoted to designing effective architectures to learn feature interactions. However, we observe that the practical performance of those designs can vary from dataset to dataset, even when the order of interactions claimed to be captured is the same. That indicates different designs may have different advantages and the interactions captured by them have non-overlapping information. Motivated by this observation, we propose DHEN - a deep and hierarchical ensemble architecture that can leverage strengths of heterogeneous interaction modules and learn a hierarchy of the interactions under different orders. To overcome the challenge brought by DHEN's deeper and multi-layer structure in training, we propose a novel co-designed training system that can further improve the training efficiency of DHEN. Experiments of DHEN on large-scale dataset from CTR prediction tasks attained 0.27\% improvement on the Normalized Entropy (NE) of prediction and 1.2x better training throughput than state-of-the-art baseline, demonstrating their effectiveness in practice.
\end{abstract}
\maketitle
\section{Introduction}
Online advertising has rapidly grown to be a multi-billion business: for the US alone, it has reached \$284.3 billion in the fiscal year 2021, with a 25\% growth compared to the fiscal year 2020 \cite{adgate_2021}. Predicting the probability that a user clicks on a particular advertisement (a.k.a. click-through rate prediction) has played an important role in online advertising. This is because the performance of the click-through rate (CTR) prediction has an impact on the user satisfaction of the business providers. 

Due to the importance of CTR prediction, tremendous efforts have been invested in improving the prediction model performance in both academia and industry \cite{mcmahan2013ad, he2014practical}. In the early stage, logistic regression (LR) \cite{mcmahan2013ad} was used to model the relationship between features and label. Unfortunately, its assumption on linearity prevents capturing nonlinear input-output relationships, limiting its capability and applicability in complex scenarios. That limitation was partially resolved by the use of decision tree (DT) in \cite{he2014practical}, where input features are first non-linearly transformed by DT. However, the success of both the LR and the DT requires exhaustive efforts on feature engineering, which in reality is resource-intensive as there is an extensive number of raw features along with feature crossings (interactions). To tackle the challenge, \cite{rendle2010factorization} proposes a factorization machine (FMs) that can capture second-order feature interaction through the inner product of latent embeddings. In the meantime, the pre-defined shallow structure of FMs limits its expressive power. Extensions of FM to more expressive formats, such as HOFMs \cite{blondel2016higher}, FFMs \cite{juan2016field,juan2017field} and AFM \cite{xiao2017attentional}, suffer from over-fitting along with undesirably high computation cost. All the aforementioned limitations can be properly handled by deep learning based models: the use of deep hidden layers and nonlinear activation functions enables capturing of nonlinear high-degree feature interactions in an end-to-end manner, relieving ranking engineers of the burden of exhaustive manual feature engineering and poor expressiveness imposed by shallow models. %predefined formula.

Since 2016, various deep models have been deployed by the business providers to serve their ads ranking services, including Wide\&Deep~\cite{cheng2016wide} (Google Play), DeepFM~\cite{guo2017deepfm} (Huawei AppGallery), DIN~\cite{zhou2018deep} (Taobao), and FiBiNET~\cite{huang2019fibinet} (Weibo). Most of those deep models consist of two primary components: feature embedding learning and feature interaction modeling. Feature embeddings are learned via mapping categorical features into embedding vectors. Feature interactions are learned by utilizing functions to model the relationship among the embeddings. Multiple studies \cite{cheng2016wide, wang2017deep, he2017neural, lian2018xdeepfm, song2019autoint, liu2020autofis, lang2021architecture} have shown that a better design of the interaction part can lift the prediction accuracy significantly over real-world applications. This has motivated a variety of studies that evolves from capturing low-order interactions towards high-order ones \cite{zhang2016deep, qu2016product, he2017neural, guo2017deepfm, wang2021dcn, lian2018xdeepfm, yan2020xdeepint, yan2020xdeepint, choi2016retain, park2018multimodal,lu2016hierarchical, li2020interpretable, shalev2017failures, liu2020autofis, lang2021architecture, zhu2021aim}. 

Although several prior studies claim their design of feature interaction modules can capture high-order interactions, we notice their practical performance ranking can vary from dataset to dataset, even when they claim to capture the same degree of interaction. This indicates that different interaction modules intended to capture the same order of interaction have different strengths over different datasets, and moreover the root cause might be the information captured by them isn't overlapping. In the meantime, as shown by the experimental parts in those studies, the performance improvement obtained through learning higher-order interaction (often realized by stacking more interaction layers) sometimes has a negative effect (Figure 3 in DCN \cite{wang2017deep}, Figure 7(a) in xDeepFM \cite{lian2018xdeepfm}, Figure 4 in InterHAt \cite{li2020interpretable}, Table 2 in xDeepInt \cite{yan2020xdeepint}, and Figure 3(a) in GIN \cite{lang2021architecture}, to name a few). This is counter-intuitive as according to theory, capturing higher-order interaction should lead to better or at least neutral performance. We hypothesize that this issue originates from the use of a homogeneous interaction module that limits the types of interactions that can be captured. This observation signals the importance of having heterogeneous interaction modules in the model. Motivated by this, we propose DHEN: A \underline{D}eep and \underline{H}ierarchical \underline{E}nsemble \underline{N}etwork with a layered structure. In a DHEN layer, there is a collection of heterogeneous interaction modules and ensemble components. The collection of heterogeneous interaction modules can complement the non-overlapping information that different interaction modules can capture. The ensemble component captures the \emph{correlation} between heterogeneous modules. As such, through recursively stacking DHEN layers, the model learns a hierarchy of the interactions of different orders and captures the correlation of heterogeneous modules, which we find empirically plays an important role in attaining better performance.

\vspace{1mm}
\noindent The main contributions of this paper can be summarized as follows:
\begin{itemize}[leftmargin=*]
    \item We design a novel architecture called Deep and Hierarchical Ensemble Network (DHEN) based on the observations that different interaction modules have different strengths over distinct datasets. Through recursively stacking interaction and ensemble layers, DHEN can learn a hierarchy of the interactions of different orders learned by heterogeneous modules. 
    \item Compared to previous CTR prediction models, DHEN's deeper, multi-layer structure increases training complexity, posing a challenge to practical training. We proposed a series of mechanisms to improve DHEN training performance, including a new distributed training paradigm called Hybrid Sharded Data Parallel that achieves up to 1.2x better throughput than state-of-the-art fully sharded data parallel to support efficient training for large DHEN models.
    \item A comprehensive evaluation is conducted for DHEN on a large-scale dataset from CTR prediction tasks, demonstrating up to 0.27\% Normalized Entropy (NE) gain over the state-of-the-art AdvancedDLRM baseline. %Although DHEN is demonstrated by experiments over datasets with clicks as positive labels, the architecture and idea behind is label agnostic and can be applied to any ranking problems. 
\end{itemize}

The rest of the paper is organized as follows. Section \ref{architecture} and Section~\ref{system} illustrate the technical details of DHEN modeling and co-designed training systems, respectively. Section~\ref{experiments} provides the empirical evaluations of DHEN over the large-scale dataset. Section~\ref{related_work} briefly introduces related works of CTR prediction and feature interaction modeling. Finally, Section~\ref{conclusion} draws the conclusions and discusses the future research directions.

\begin{figure}[h]
	\centering
	\footnotesize
	\includegraphics[width=\linewidth]{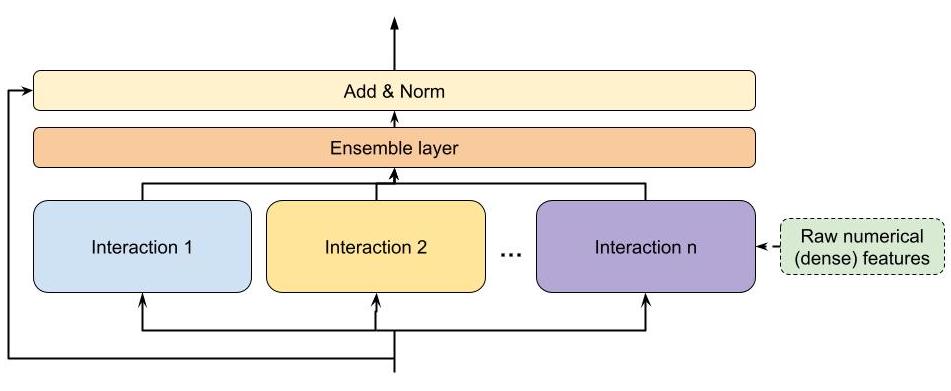}
	\caption{An general hierarchical ensemble building block in DHEN.} 
	\label{fig:general DHEN block}
\end{figure}
\section{Architecture}\label{architecture}

\begin{figure*}[ht]
\begin{center}
	\centering
	\footnotesize
	\includegraphics[scale=0.38]{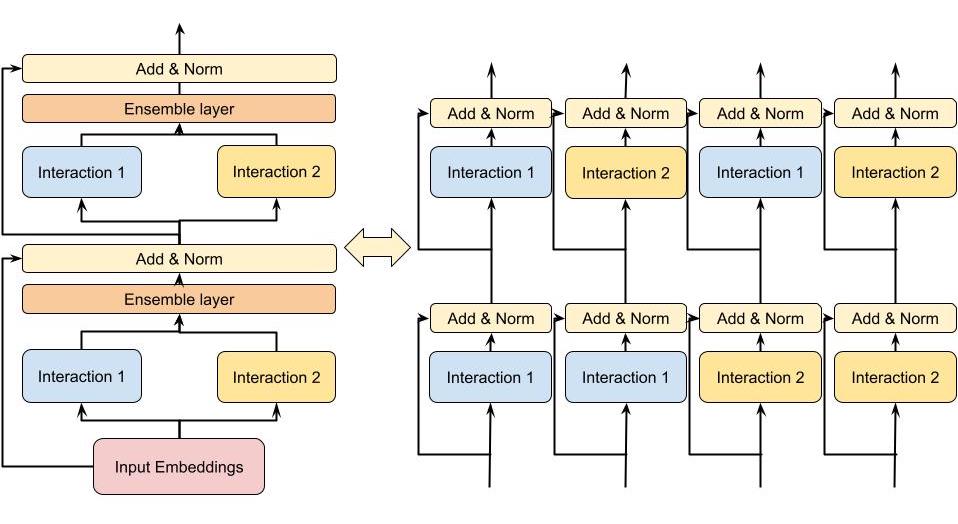}
\end{center}
\caption{A Two-layer two-module hierarchical ensemble (left) and its expanded details (right). A general DHEN can be expressed as a mixture of multiple high-order interactions. We omit the potential dense feature input for the interaction modules in this figure for clarity.}  %\arvind{minor nit - it would be good to have fig 1 before table 1.}}
\label{fig: two module Hierarchical ensemble}
\end{figure*}

Most state-of-the-art high-performance model architectures across multiple prediction tasks use a deep stacking structure (e.g., Resnet, Transformers, Metaformer \cite{he2016deep, vaswani2017attention, yu2021metaformer}). Here, the deep stacking structures are usually composed of a repeating block containing the same interaction module. For example, the transformers use self-attention blocks for stacking, and ResNet uses convolution blocks. Each block consumes the output of the previous block or the original embedding tokens as input. 
Structurally, DHEN follows this overall stacking strategy but at the same time builds a novel hierarchical ensemble framework to capture the correlations of multiple interaction modules. Figure \ref{fig:general DHEN block} shows a general DHEN building block in which an ensemble of multiple Interaction modules resides. Note that the raw numerical (dense) features can be part of the input to any modules for ensembling in every layer.

\subsection{Feature Processing Layer}
\begin{figure}[h]
	\centering
	\footnotesize
	\includegraphics[width=.9\linewidth]{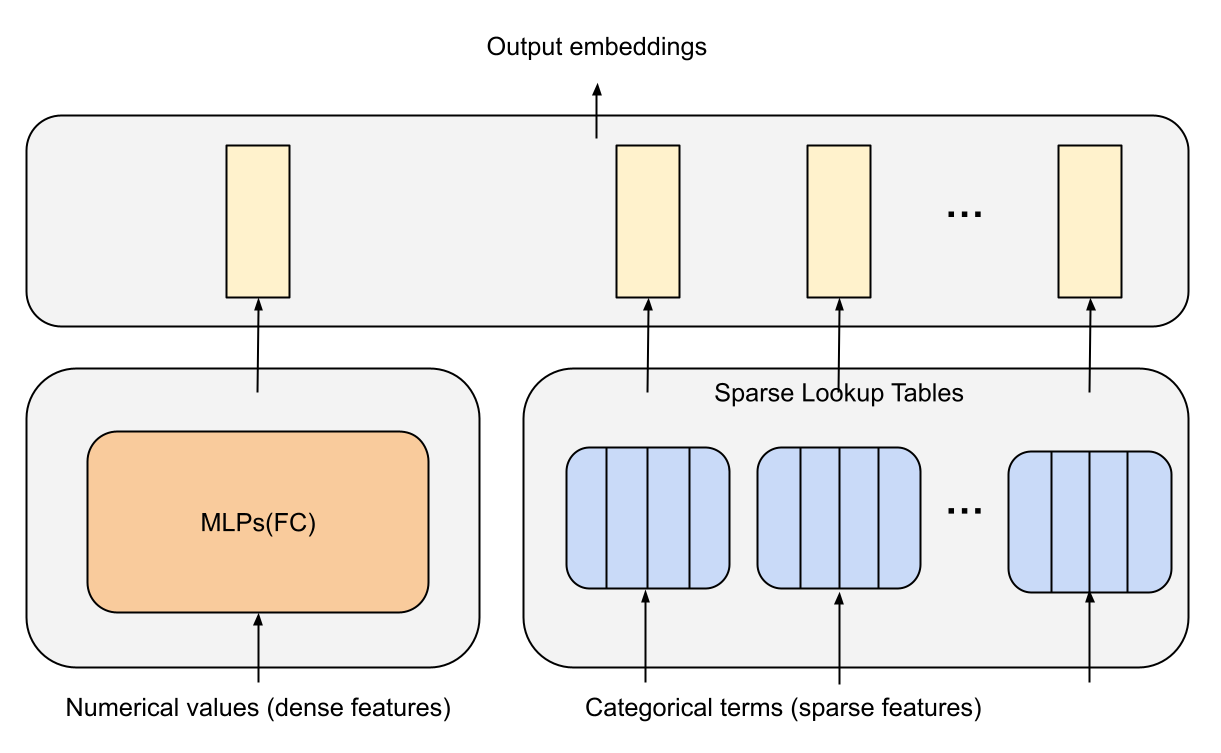}
	\caption{The feature processing layer in DHEN}  %\arvind{minor nit - it would be good to have fig 1 before table 1.}}
	\label{fig:feature processing layer}
\end{figure}

In CTR prediction tasks, the feature inputs usually contain discrete categorical terms (sparse features) and numerical values (dense features) \cite{naumov2019deep, covington2016deep, guo2017deepfm, shan2016deep}. In this work, we use the same feature processing layer in DLRM \cite{naumov2019deep}, which is shown in the Figure \ref{fig:feature processing layer}. The sparse lookup tables map the categorical terms to a list of numerical embeddings. Specifically, each categorical term is assigned a trainable d-dimensional vector as its feature representation. On the other hand, the numerical values are processed by dense layers. Dense layers compose of several Multi-layer Perceptions (MLPs) from which an output of a d-dimensional vector is computed. After a concatenation of the output from sparse lookup table and dense layer, the final output of the feature processing layer $X_0 \in \mathbb{R}^{d \times m}$ can be expressed as $X_0 = (x^{1}_0, x^{2}_0, ..., x^{m}_0)$, where $m$ is the number of the output embeddings and $d$ is the embedding dimension.

\subsection{Hierarchical Ensemble}
We propose a novel hierarchical ensemble framework that includes multiple types of interaction modules and their correlations. Conceptually, a deep hierarchical ensemble network can be described as a deep, fully connected interaction module network, which is analogous to a deep neural network with fully connected neurons.  

The hierarchical ensemble mechanism provides a framework for an ensemble of multiple interaction modules with a layer and stacking multiple layers. The input of each layer is represented as a list of embeddings, denoted by $X_n \in \mathbb{R}^{d \times m}$. Note that input of the first layer is the list of embeddings $X_0$ from the feature processing layer, and the $n$ here denotes the n-$th$ stacked layers. Formally, the output of each layer is:
\begin{equation}
    Y = Norm(Ensemble_{i=1}^{k}Interaction_i(X_{n}) + ShortCut(X_{n}))
\end{equation}
\begin{equation}
ShortCut(X_{n}) = 
    \begin{cases}
        X_{n}, &\text{if $len(X_{n}) == len(Y)$} \\
        W_{n}X_{n}, &\text{if $len(X_{n}) != len(Y)$}
    \end{cases}
\end{equation}
Where $Norm()$ here denotes a normalization method such as Layer Normalization \cite{ba2016layer}. The $Ensemble_{i=1}^{k}()$ denotes the ensemble method for $Interaction_i, Interaction_2, ..., Interaction_k$, it can be concatenation, sum, or weighed sum, etc. The $ShortCut()$ function here serves both as residual and a way to match dimensions. If the dimension of the ensemble output and the last layer input $X_{n-1}$ don't match, a linear projection $W_{n} \in \mathbb{R}^{len(X_{n}) \times len(Y)}$ will be applied to match the dimensions. Finally, the ensemble result and the shortcut are combined through an element-wise sum, and the output $Y$ will be the input of the next layer $X_n$. 

The main goal of hierarchical ensemble is to capture the correlation of the interaction modules. Figure \ref{fig: two module Hierarchical ensemble} (left) shows a two-layer two-module hierarchical ensemble component. Unlike traditional stacking model structures that only capture one type of higher-order interaction, hierarchical ensembles can capture a mixture of high-order interactions. As shown in the right part of Figure \ref{fig: two module Hierarchical ensemble}, the design of DHEN enjoys the benefit of capturing complicated higher-order interactions between the features by letting each module consume outputs of various interaction modules (e.g., $Interaction_1(Interaction_1), Interaction_1(Interaction_2),\\ Interaction_2(Interaction_1)$, and $Interaction_2(Interaction_2)$). Thus, this mixture of interaction modules can leverage multiple feature interaction types to achieve better model prediction accuracy. 

\subsection{Interaction Modules} \label{inter_modules}
We applied five types of interaction modules in our model: AdvancedDLRM, self-attention, Linear, Deep Cross Net, and Convolution. In practice, the interaction modules that can be included in DHEN are not limited to the five options above. Also, note that if an interaction module outputs a single tensor $v \in \mathbb{R}^{1 \times h}$, such as a tensor output from MLPs, we will apply a $W_m \in \mathbb{R}^{h \times (d*l)}$ to map it to a list of embeddings with dimension $d$. The $l$ here represents the number of output embeddings from the interaction module. We now provide a brief introduction of these interaction modules used in DHEN.
\subsubsection{AdvancedDLRM}
We use a DLRM style interaction module to capture the feature interactions. We call it AdvancedDLRM in our following experiments. Given an input of embeddings $X_n$, a new list of embeddings $u \in \mathbb{R}^{d \times l}$ is output by the AdvancedDLRM:
\begin{equation}
    u = W_m \cdot AdvancedDLRM(X_n)
\end{equation}

\subsubsection{Self-attention}
We consider self-attention, widely used in transformer networks for its superior performance in text understanding, as an interaction module in this paper. Self-attention was also adopted in CTR prediction tasks before~\cite{li2020interpretable}. A typical transformer includes multiple stacking encoder/decoder layers, with the self-attention mechanism at their core. In this paper, given an input of embeddings $X_n$, we apply a transformer encoder layer as:
\begin{equation}
    u = W \cdot TransformerEncoderLayer(X_n)
\end{equation}
where $W \in \mathbb{R}^{m \times l}$ is used to match and unify the output dimensions from all interaction modules.

\subsubsection{Convolution}
Convolution layers are widely used in computer vision tasks. It was also adopted in NLP, and CTR tasks \cite{liu2019feature, gulati2020conformer}. In this paper, we adopt convolution as one of the interaction modules. Given an input of embeddings $X_n$, a list of embeddings $u$ is obtained from:
\begin{equation}
    u = W \cdot Conv2d(X_n)
\end{equation}
Similarly, $W \in \mathbb{R}^{m \times l}$ is also used to match and unify the output dimensions from all interaction modules.

\subsubsection{Linear}
Linear layers is one of the most straightforward modules to capture the raw information from the original feature embeddings. In this paper, we use the linear layer as one of the interaction modules to condense the information in each layer. Given an input of embeddings $X_n$, a list of embeddings $u$ is obtained from:
\begin{equation}
    u = W \cdot (X_n)
\end{equation}
Where $W \in \mathbb{R}^{m \times l}$ here is used as a linear module weight to match the dimensions. 
\subsubsection{Deep Cross Net}
Deep Cross Net (DCN) is a widely used feature interaction Module in CTR prediction tasks \cite{wang2017deep}. It introduces a cross network that is efficient in learning certain bounded-degree feature interactions. In this paper, we adopt the DCN module as one of the interaction modules in each layer. Given an input of embeddings $X_n$, a list of embeddings $u$ is obtained from:
\begin{equation}
    u = W \cdot (X_n \cdot X^T_n) + b
\end{equation}
Where $W$ and $b$ denote the weight and bias metric in the DCN modules. We omit the skip connection process from the original paper in the equation above because we already used skip connection to flow information across the stacked layers.

\section{Training System}\label{system}
Intuitively, the depth of the stacked hierarchical ensemble layers contributes to the expressiveness of DHEN, but they also create a challenge in practical training of DHEN. This section highlights how we enable efficient and scalable training of DHEN in our cluster.

\subsection{Training Strategy}
\label{label:training_strategy}
Each DHEN training sample contains both categorical and numerical features, and all features must be converted to dense representations before they can be consumed by the DHEN layers. One immediate challenge of this process is the sheer size and complexity of DHEN completely overshadows the capacity of a single standard datacenter server, typically hosting 8 GPUs with an aggregate high bandwidth memory of a few hundred gigabytes and compute capability of up to tens of petaflops. Thus a single server is drastically underpowered compared to the typical number of parameters (up to trillions) and flops needed (up to giga flops) to train a single sample in DHEN.

To allow efficient training of DHEN, we take advantage of the ZionEX fully synchronous training system detailed in~\cite{mudigere2021softwarehardware}. At a high level, the ZionEX system groups 16 hosts into a "supernode``, called a pod, which contains 128 A100 GPUs (8 per host) with a total HBM capacity of 5TB and 40PF/s BF16 compute capability. Within a host, each GPU is connected through NVLink, and each host in a pod is then connected with a high bandwidth network of up to 200GB/s, shared with 8 GPUs.

With ZionEX, we solve the compute and memory capacity issue with the following distributed training strategy. We first distribute the embedding tables across a pod. To provide better load balancing and deal with oversized embedding tables, instead of placing whole tables to different GPUs, we proactively slice oversized embedding tables into equal column shards, and place these columns based on an empirical cost function. Our cost function captures both compute and communication overhead of such placement. We distribute the table shards in a load balanced manner dictated by the cost function, using the LPT~\cite{Longestp74:online} approximate set partition algorithm. On the other hand, for dense modules including the DHEN layers, we replicate them on each GPU and train them in a data parallel (DP) fashion. This choice is based on the observation that the activation of the dense DHEN layers can be much larger than the weights themselves, and thus synchronizing the weights has a lower  cost than sending the activations through the network. This training strategy thus induces a hybrid training paradigm, where each batch starts with DP, enters model parallelism for distributed embedding lookup, and ends with DP for the dense layers. 

Training dense modules using DP imposes a parameter size ceiling equal to per-GPU HBM capacity on the stacked DHEN layers, which hinders our exploration into the limits of DHEN scalability. To solve this issue, we use fully sharded data parallel (FSDP~\cite{10.1145/3394486.3406703, FairScale2021}) to remove memory redundancy in traditional data parallelism by further sharding weights to different GPUs, activation checkpointing to trade more compute for less peak memory usage~\cite{chen2016training}, cpu offloading~\cite{rajbhandari2021zeroinfinity} to further reduce GPU memory usage by aggressively storing parameters and gradients to CPU, and bring them back to GPU right before needed. Since all of these techniques hurt training efficiency, we carefully tune the system to turn on a minimum set of them for our training needs based on the number of DHEN layers.

\begin{figure}[t]
	\centering
	\footnotesize
	\includegraphics[width=\linewidth]{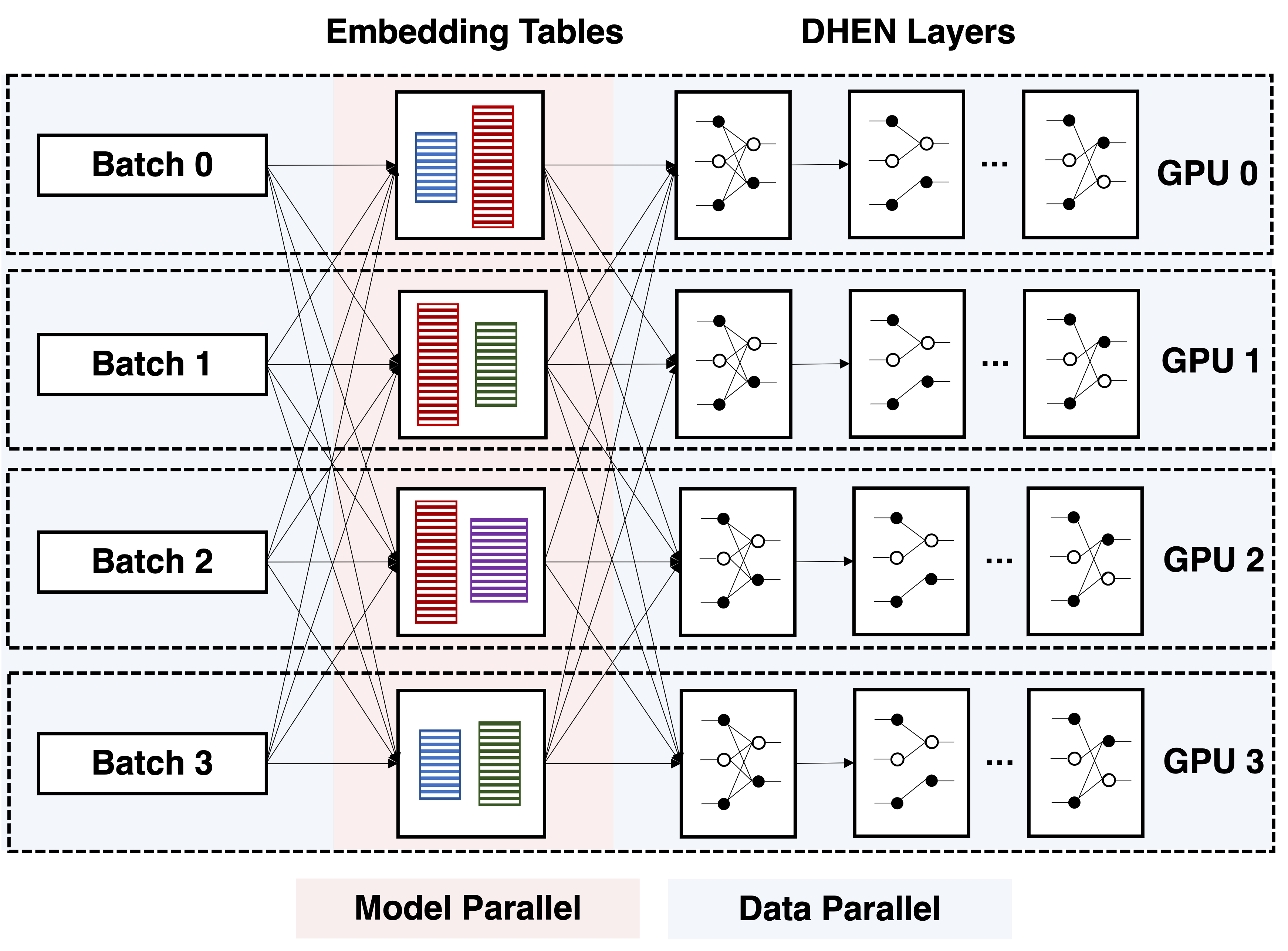}
	\caption{Training strategy for DHEN (4 GPUs shown).} %. The fast nodes are up to 1.2x faster than the slow ones.} %\arvind{minor nit - it would be good to have fig 1 before table 1.}}
	\label{fig:dhen_training}
\end{figure}

The described training strategy of DHEN is summarized in Figure~\ref{fig:dhen_training}.

\subsection{Training Optimizations}
We provide additional optimizations to further speedup training.

\subsubsection{Common Optimizations}
We enable a set of widely-used optimizations including large batch training~\cite{goyal2018accurate} to reduce synchronization frequency, FP16 embedding with stochastic rounding, BF16 optimizer, and quantized all to all and allreduce collectives~\cite{zhang2018training, yang2020training} to further reduce memory footprint, help with numeric stability, leverage specialized accelerator hardware such as Tensor Core~\cite{tensorcore} and to reduce communication overhead.

\subsubsection{Hybrid Sharded Data Parallel}
To find the best DHEN configuration within the training cost budget, we must probe both the configuration of token mixers and the number of layers. In practice, we must frequently experiment with a candidate DHEN model whose memory footprint just exceeds the per-GPU HBM capacity (we call these models ``embarrassingly-sized''), which renders data parallel infeasible and necessitates FSDP. However, in these cases, we find that applying FSDP directly does not result in optimal efficiency, especially when training at a production scale (hundreds of GPUs). This is because the allgather operations that brings different shards of weights from each GPU, required on the critical path of both forward and backward passes, can take long to finish as (1) it needs to communicate with all GPUs in the cluster, and (2) each shard would be too small to efficiently leverage network bandwidth. Although techniques such as prefetching help alleviate this problem, they add memory pressure to the system, and thus defeats the purpose of using FSDP in the first place. 

\begin{figure}[t]
	\centering
	\footnotesize
	\includegraphics[width=.9\linewidth]{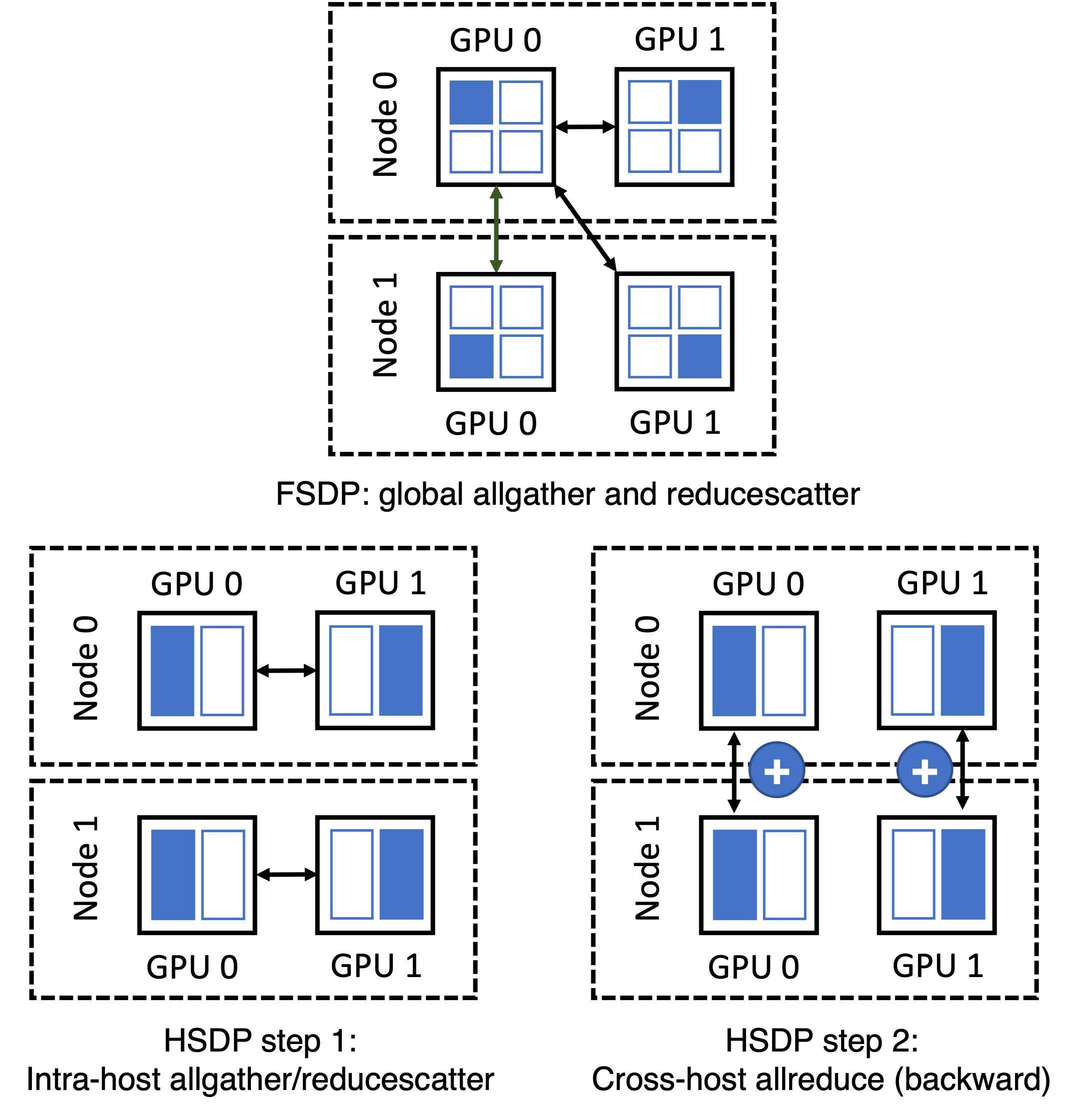}
	\caption{Comparison between FSDP (top) vs HSDP (bottom). Shown with 2 hosts with 2 GPUs each. We train large DHENs using FSDP, and medium-sized DHENs using HSDP.} %. The fast nodes are up to 1.2x faster than the slow ones.} %\arvind{minor nit - it would be good to have fig 1 before table 1.}}
	\label{fig:fsdp_vs_hsdp}
\end{figure}

To that end, we propose a novel training paradigm for these embarrassingly-sized networks, called \textit{hybrid sharded data parallel} (HSDP), co-designed with the DHEN model for our training cluster. HSDP recognizes the 24X bandwidth difference between the GPU interconnect (NVLink, 600GB/s) and host interconnect (RoCEv2, 25GB/s), and work as follows: (1) HSDP shards the entire model within a single host, so that the reducescatter operation in the backward pass, and the allgather operation in both the forward and backward pass in FSDP operate completely within a host; (2) when the reducescatter operation finishes in the backward pass, we hook concurrent allreduce operations, each one involving all GPUs on each host with the same local host ID, to compute an average gradient for its local shard, in an asynchronous manner to avoid blocking computation of the backward pass. (3) at last, we register a backward hook to wait on the handles of the pending allreduce operations, so the semantic is left intact in HSDP compared to FSDP and DDP from a training perspective. We contrast FSDP and our HSDP in Figure~\ref{fig:fsdp_vs_hsdp}.

Qualitatively, compared to FSDP, HSDP provides a few benefits: (1) the latency of allgather operation on the critical pass of both forward and backward pass is significantly reduced because and it can enjoy the high speed NVLink interconnect without needing to communicate cross host through the slower RoCE link; (2) the scale at which the communication collectives operates scales with the number of host, instead of number of GPUs, promoting efficiency. The tradeoff for HSDP, however, is that it supports models at most 8x (number of GPUs per host) the size supported by pure data parallel, and incurs an 1.125x communication overhead in terms of bytes on wire due to the added allreduce operation. These are acceptable for these embarrassingly-sized models, because the allreduce overhead can be hidden with computation of next layers and allreduce is a highly optimized operator.

\begin{table*}[ht]
\centering
\begin{tabular}[width=.9\linewidth]{cccccc}
    \hline
        Model id & Interaction type(s) & NE diff @10B examples & NE diff @20B examples & NE diff @35B examples & $N$ \\ 
    \hline
        1 & DCN(baseline) & NA & NA & NA & 5\\
    \hline
        2 & Self-attention & 0.036\% & 0.026\% & -1.044\% & 5\\
    \hline
        3 & CNN & -1.441\% & -1.535\% & -1.534\% & 5\\
    \hline
        4 & Linear & \textbf{-1.461\%} & \textbf{-1.546\%} & -1.538\% & 5\\
    \hline  
        5 & DCN + Linear & -0.002\% & -0.004\% & -0.004\% & 5\\
    \hline
        6 & Self-attetion + Linear & -1.363\% & -1.537\% & \textbf{-1.576\%} & 5\\
    \hline
        7 & Self-attetion + CNN & 0.024\% & -1.270\% & -1.508\% & 5 \\
    \hline
\end{tabular}
\caption{Model variations on the hierarchical ensemble architecture. The hierarchical ensemble of Self-attention and Linear interaction modules shows the best prediction accuracy, and significantly helps the learning of self-attention interaction converge. Here, NE denotes the Normalized Entropy \cite{he2014practical}}
\label{model_variants_results}
\end{table*}

\begin{table*}[ht]
\centering
\begin{tabular}[width=.9\linewidth]{cccccc}
    \hline
        Model id & Interaction type(s) & $N$ & NE diff & NE diff & NE diff \\ 
        ~ & ~ & ~ & @5B examples & @15B examples & @25B examples \\
    \hline
        1 & AdvancedDLRM(baseline) & 1 & NA & NA & NA \\
    \hline
        2 & AdvancedDLRM + Linear & 2 & -0.0315\% & -0.134\% & -0.176\% \\
    \hline
        3 & AdvancedDLRM + Linear & 4 & -0.071\% & -0.197\% &  -0.255\%\\
    \hline
        4 & AdvancedDLRM + Linear & 8 & \textbf{-0.068\%} & \textbf{-0.208\%} & \textbf{-0.273\%} \\
    \hline  
\end{tabular}
\caption{DHEN performance vs. Industrial AdvancedDLRM model. Deeper DHEN captures higher order interactions and shows better prediction accuracy.}.
\label{p_model_r}
\end{table*}

\begin{figure*}[h]
    %\begin{minipage}[t]{.3\linewidth}
    %\vspace{0pt}
    %\centering
    %\includegraphics[width=5.5cm]{figures/DHEN vs MoE.png}
    %\label{fig: MoE vs. DHEN}
    %\end{minipage}%
    \begin{minipage}[t]{.7\linewidth}
    \vspace{0pt}
    \centering
    \begin{tabular}{cccc}
        \hline
            Model id & Scaling method & Training FLOPs & NE diff @50B example \\
        \hline
            1 & AdvancedDLRM (baseline) & 0.06G & NA \\
        \hline
            2 & 4 expert MoE & 1.3G & -0.06\% \\
        \hline
            3 & 2 layer DHEN & 1.44G & \textbf{-0.11\%} \\
        \hline
            4 & 8 expert MoE & 3.3G & -0.09\% \\
        \hline
            5 & 4 layer DHEN & 3G & \textbf{-0.21\%} \\
        \hline
            6 & 16 expert MoE & 6G & -0.10\%\\
        \hline
            7 & 6 layer DHEN & 4.6G & \textbf{-0.26\%} \\
        \hline  
    \end{tabular}
    \end{minipage}
    \captionof{table}{DHEN achieves better scaling efficiency than MLP scaling by MoE.}
    \label{moe vs dhen}
\end{figure*}

\section{Experiments}\label{experiments}
In this section, we conduct experiments to evaluate the effectiveness of the Deep Hierarchical Ensemble Network for CTR prediction tasks. Our experiments has the following objectives: (1) assess the effectiveness of hierarchical ensemble and identify a good set of interaction modules for ensemble; (2) evaluate the end-to-end accuracy gain of DHEN compared to a state-of-the-art DLRM model, and demonstrate the effectiveness of our system-level optimizations on the training throughput of DHEN.

\subsection{Experiment Setup}
We use an industrial dataset for all experiments. As previously mentioned, the in-house AdvancedDLRM model was considered as one of the interaction modules. We omit the detailed descriptions for the AdvancedDLRM and the dataset settings for simplicity. Training hyper-parameters, including learning rate and optimizer configurations are the same for all the experiments. All models are trained with hundreds of sparse (categorical) features and thousands of dense (numerical) features. The full-sync training scheme ensures both model performance and training throughput can be reproduced \cite{mudigere2021software}. We use Normalized Entropy loss to evaluate the CTR prediction accuracy \cite{he2014practical}. 

\subsection{Model Variation with Different Interaction Modules}
The purpose of this section is to understand the behavior of DHEN with commonly used modules mentioned in Section \ref{inter_modules}. Specifically, we used the four interaction modules (DCN, Self-attention, CNN, and Linear) to test the performance of hierarchical ensemble (we compare with AdvancedDLRM in the next section in depth). In each experiment, we include different types of interaction modules in each layer. Table \ref{model_variants_results} shows the performance of different model settings, where $N$ denotes the number of stacked layers. To facilitate comparison, we use DCN as baseline and use relative Normalized Entropy loss difference at different training steps (training examples) to evaluate the model performance.

From Table \ref{model_variants_results}, we observed that the model (4) with linear interaction module shows the best performance among the models with single interaction module per layer, and the model (2) with self-attention interaction module needs large training data to converge. We further observed that, after the hierarchical ensemble of the self-attention and linear interaction module, model (6) starts to show the best performance among all the variants. This result also suggested that the hierarchical ensemble architecture captured the correlation of the self-attention and linear interaction module, and significantly helped the self-attention interaction module to converge. Further, we noticed that it's not always beneficial to have more interaction modules for ensemble in each layer: for example, model (5) has an hierarchical ensemble of DCN and linear layers, and its performance is better than stacking the DCN layer but worse than using linear layer alone; a similar situation can be found for model (7) with an hierarchical ensemble of self-attention and CNN interaction modules. These observations confirm our initial hypothesis that different interaction modules capture non-overlapping interactions, and hence the hierarchical ensemble mechanism is key to the performance of DHEN. 

\subsection{Experiments On Industrial AdvancedDLRM}
\subsubsection{Model Prediction Performance}
After examining the effectiveness of DHEN with various interaction modules, we take one step further to evaluate its effectiveness on top of the industrial module, AdvancedDLRM. Specifically, we leverage the state-of-the-art AdvancedDLRM \cite{naumov2019deep} and the Linear module as two interaction modules for hierarchical ensemble. Table \ref{p_model_r} shows the DHEN performance compared to the AdvancedDLRM model.

We trained 4 DHEN models with a hierarchical ensemble of AdvancedDLRM and Linear interaction modules in each layer. Each of them has $N$ layers. From Table \ref{p_model_r}, we observed that all the DHEN models outperform the industrial AdvancedDLRM model, and deeper DHEN models achieve larger Normalized Entropy improvement. In both cases, we observed that the gain keeps enlarging as more training examples are processed. This suggests that both higher order interactions and the correlations of various interactions play an important role in the CTR prediction tasks, and the enlarging Normalized Entropy improvement with larger datasets indicates DHEN performance is both consistent and generalizable. 

\subsubsection{Scaling Efficiency}
To evaluate DHEN model scaling efficiency, we summarize our results with a comparison between two scaling methods using MoE \cite{shazeer2017outrageously} and stacking DHEN layers, in Table \ref{moe vs dhen}. Both scaling methods were implemented based on the same industrial AdvancedDLRM \cite{naumov2019deep}. We use the MoE architecture to scale up the MLP layers in AdvancedDLRM while using an ensemble of AdvancedDLRM and Linear modules to stack DHEN layers. The results show that DHEN consistently beats AdvancedDLRM MoE with similar training complexity (measured in FLOPs) in terms of accuracy, and thus stacking DHEN layers stands out as an effective mechanism for scaling with superior return of investment.

\subsubsection{Training Throughput}
This section assesses the throughput of training DHEN model on a 256-GPU cluster based on the ZionEX design introduced in Section~\ref{label:training_strategy}. We start by demonstrating the effectiveness of our system level optimizations with a 8-layer DHEN model trained using DP. Overall, we see 1.08x end to end speedup from applying FP16 embedding, AMP, quantized BF16 allreduce and all to all collectives. The improvement comes from the reduced exposed communication latency with the quantized collectives, and the remaining bottleneck lies in the optimizer cost and the all to all collectives calls that cannot be fully overlapped with the compute of dense layers. 

\begin{figure}[h]
	\centering
	\footnotesize
	\includegraphics[width=.9\linewidth]{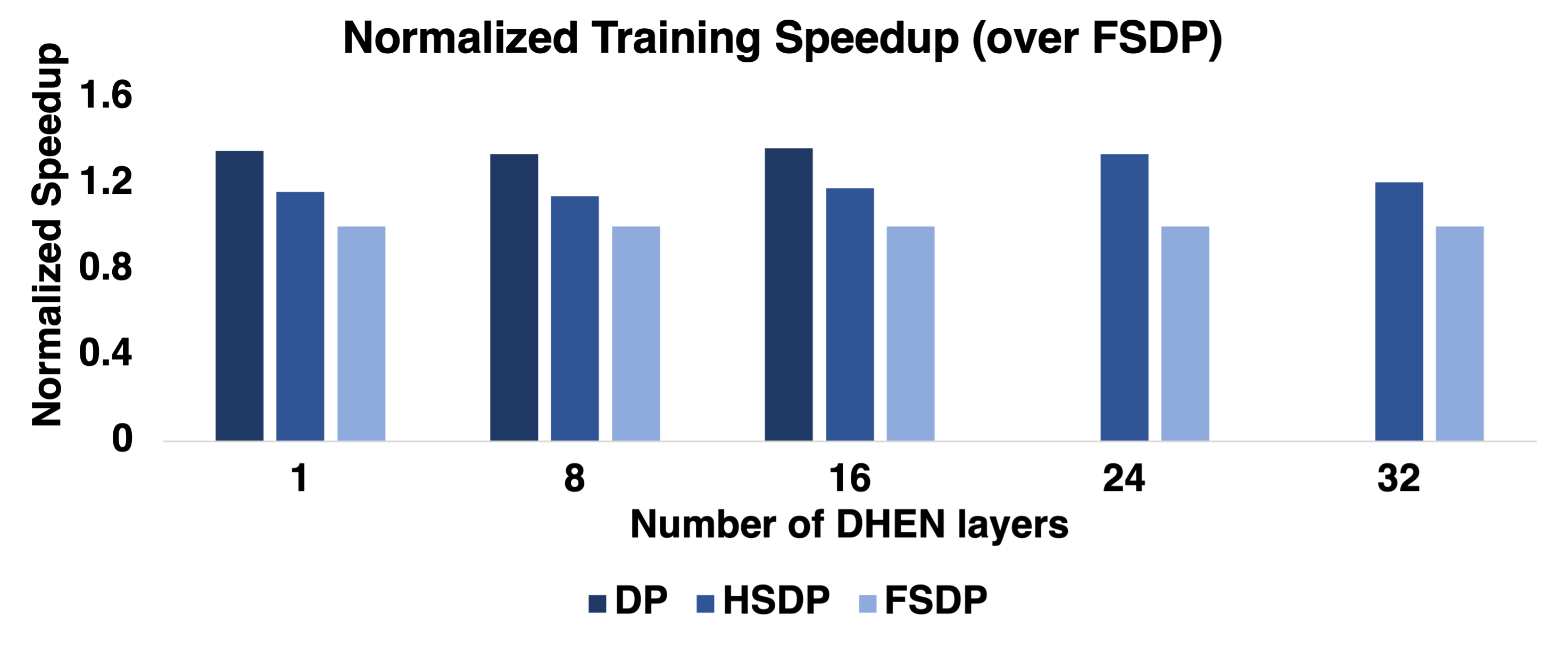}
	\caption{HSDP supports larger model sizes compared to DP and higher training throughput compared to FSDP.} 
	\label{fig:hsdp_vs_fsdp_vs_ddp}
\end{figure}

While DHEN scales well with more layers, training deeper DHEN models require paradigm shift as DP can only support up to 22 layers of DHEN in our cluster. We now quantify how our proposed training paradigm codesigned with DHEN, HSDP, can help bridge both the throughput gap of training with FSDP compared to DP, as well as close the memory gap of training DHEN with DP. We conduct training throughput experiments on various DHEN models with different the number of DHEN layers. We train each model using all three paradigms and compare the resulting throughput. As shown in Figure~\ref{fig:hsdp_vs_fsdp_vs_ddp}, DP performs well up to 16 layers, and further increase in layers results in errors; on the other hand, both FSDP and HSDP support larger layer counts, and HSDP consistently outperforms FSDP by up to 1.2x. Our trace analysis shows the throughput gain from HSDP indeed comes from drastically reduced allgather latency on the training critical path, thanks to the fast NVLink connection within the single host.

\section{Related Work and Discussion}\label{related_work}
\balance
\noindent\textbf{CTR Prediction Models.}
In the early stage, logistic regression (LR) \cite{mcmahan2013ad} was adopted as the model for CTR prediction task. Unfortunately, it is a linear model and thus unable to capture the nonlinear relationship between the features and the outcome. That limitation was later resolved by decision tree (DT) \cite{he2014practical} based feature transformation. Yet, both LR and DT require exhaustive efforts on feature engineering to succeed, which is difficult especially when the number of raw features is large and high-order feature interaction is present. To attack the problem, factorization machine (FMs) \cite{rendle2010factorization} was proposed to model second-order feature interactions through the inner product of their latent embeddings. However, the pre-defined shallow structure of FMs limits its expressiveness. Later, FMs was extended to higher order in \cite{blondel2016higher}, but due to its need to enumerate all high-order interactions without differentiating their importance, the large number of parameters to be learned for their latent vectors causes impractically high computational cost in the real-world systems. FFMs \cite{juan2016field,juan2017field} extends FM, by assigning each feature with multiple vectors to interact with features from other fields. Despite the significant performance improvement, FFMs introduce many more parameters and suffer from over-fitting issues. AFM \cite{xiao2017attentional} extends FMs with an ``attention net'' that improves not only the performance but also interpretability. The authors argue that the feature salience provided by the attention network greatly enhances the transparency of FMs. That said, AFM can only learn up to the second-order attention-based salience due to the inherent architectural limit of FMs. Those concerns can be properly addressed by deep learning based models: the use of deep hidden layers and nonlinear activation functions enable those models to capture nonlinear high-degree feature interactions in an end-to-end manner, getting rid of the exhaustive manual efforts of feature engineering and the limitations in expressiveness by the predefined formula. Since 2016, various deep models have been deployed by the business providers on their ads ranking services, such as Wide\&Deep \cite{cheng2016wide} on Google Play, DeepFM \cite{guo2017deepfm} on Huawei AppGallery, DIN \cite{zhou2018deep} on Taobao, and FiBiNET on Weibo \cite{huang2019fibinet}. Most deep CTR prediction models consist of two primary components: feature embedding learning and feature interaction modeling. Feature embeddings are learned via mapping categorical features into low-dimensional embedding vectors. Feature interactions are learned by utilizing some functions to model the relationship among the embeddings. Multiple studies \cite{cheng2016wide, wang2017deep, he2017neural, lian2018xdeepfm, song2019autoint, liu2020autofis, lang2021architecture} have shown that a better design of the interaction part can generate significant lift on the prediction accuracy, which motivates a variety of designs from different perspectives. 

\noindent\textbf{Feature Interaction.} \label{feature_interaction}
In FNN \cite{zhang2016deep}, fully connected layers are used to learn higher-order interaction on top of the second-order interaction from pre-trained FMs. As a result, its performance is largely limited by the capability of the pre-trained FMs. PNN \cite{qu2016product} mitigates the limitation through replacing the pre-trained FMs by a product layer. NFM \cite{he2017neural} extends FMs by replacing the inner-product with a Hadamard product. Similarly like PNN and FNN, NFM stacks deep neural networks (DNNs) on top of the second-order feature interactions to model higher-order features. The major downside of FNN and PNN is that they both focus more on high-order feature interactions while capturing little low-order interactions, which are also essential for CTR prediction. To model both low-and-high-order feature interactions, Wide\& Deep \cite{cheng2016wide} proposes a hybrid network structure that combines a shallow part for artificial cross-features and a deep part for raw features. The limitation is that the input of the shallow part still relies on manual efforts of feature engineering. To mitigate that, DeepFM \cite{guo2017deepfm} imposes a FMs as the shallow part of Wide\&Deep to capture the cross-features (second-order interaction). DCN \cite{wang2017deep,wang2021dcn} replaces the shallow part of Wide\&Deep with a cross-product transformation that can efficiently capture feature interactions of bounded degrees. Unfortunately, as pointed out by \cite{lian2018xdeepfm} its output is constrained to be a very specific format and its captured interactions are solely at the bit-wise level. xDeepFM \cite{lian2018xdeepfm} improves DeepFM through introducing higher-order interactions and improves DCN through introducing replacing the cross network in DCN with compressed interaction network - a more general interaction module that can capture feature-wise interactions. It has been noted by \cite{yan2020xdeepint} that the aforementioned methods ignore the potential benefits of combining the feature-wise and bit-wise interactions or rely on deep and resource-intensive structure to realize the latter. To overcome the limitation, xDeepInt \cite{yan2020xdeepint} proposes to utilize subspace crossing between individual interaction layers to learn high-order feature-wise and bit-wise interactions recursively, dispensing with jointly-trained DNN and nonlinear activation functions. 

In recent years, explainability has attracted  more attention due to an increasing desire in reliability and security. Although more than one technique such as DCN and xDeepFM could capture high-order interactions, they suffer from poor explainability. Inspired by the success of Transformer \cite{vaswani2017attention} in mining feature relevance as well as the attention mechanism's capability in locating features greatly affecting predictions \cite{choi2016retain, park2018multimodal,lu2016hierarchical}, AutoInt \cite{song2019autoint} proposes to use multi-layer, multi-head, and self-attentive neural networks with residual connection to learn different orders of feature interactions explicitly and meanwhile offer good model explainability. Similarly like AutoInt, InterHAt \cite{li2020interpretable} proposes to capture high-order feature interactions through an attentional aggregation strategy that provides good explainability and also proves to have lower computational cost than Wide\&Deep and DCN. This is by no means to enumerate all studies on interaction modules but provide representative examples. A more comprehensive review can be found in \cite{zhang2021deep,yang2022click}. Although multiple aforementioned studies have shown the benefits of high-order interaction in prediction accuracy, a more in-depth look reveals that those high-order interactions are not equally important. In fact, useless interactions may bring unnecessary noise that impairs the learning process and leads to worse performance \cite{shalev2017failures}. However, it is challenging to identify their importance manually due to exponentially growing combination space for interactions. To automatically learn which feature interactions are essential, AutoFIS \cite{liu2020autofis} introduces a gate (in open or closed status) for each feature interaction to control whether its output should be passed to the next layer. Similarly, GIN \cite{lang2021architecture} relies on a second-stage re-training to prune unimportant feature interactions. AIM \cite{zhu2021aim} further improves AutoFIS through relaxing discrete selection of open gates to be continuous parameters that can be jointly trained with model parameters and allowing for dedicated embedding size for features of different importance. Although several above studies claim their design of interaction part can capture high-order interaction, we notice their practical performance can vary from dataset to dataset, even when their bounded degrees of the interaction are the same. That indicates, the information captured by different interaction modules still has a difference, thus demonstrating different strengths over different datasets. In the meantime, as shown by those studies, the performance improvement obtained through learning higher-order interaction (often realized by stacking more interaction layers) sometimes have an opposite effect. This motivates our study on DHEN.

In the end, it is worth mentioning the architectures closest to DHEN from prior studies are GIN \cite{lang2021architecture} and AutoInt \cite{song2019autoint}. Compared to them, our study has salient differences: although GIN and AutoInt also uses more than one interaction module (the multi-head attention structure in GIN and the multi-branch structure in AutoInt), each head is still homogeneous, which cannot complement non-overlapping information and capture the correlation of different modules as well as the hierarchy. Further, AutoInt also needs a second-stage re-training, limiting its applicability whereas DHEN follows an end-to-end manner and only needs to be trained once, making it more feasible in different applications.

\noindent\textbf{Future Work.}
DHEN provides a solid foundation that opens up greater opportunities for even more accurate CTR prediction. For instance, we can apply a dedicated gating activation to individual layers of DHEN so that different examples can harness different order of interaction in the hierarchy; we can also introduce Mixture-of-Experts structure to each layer of DHEN to allocate dedicated hierarchy for individual examples; finally, we can also enable shared and dedicated layers in DHEN for different tasks to enable DHEN for multi-task scenarios. 
\section{Concluding Remarks}\label{conclusion}
In this paper, motivated by the observation that existing interaction modules may possess different advantages on different datasets, we propose DHEN - a hierarchical ensemble architecture that can leverage strengths of heterogeneous interaction modules and learn a hierarchy of the interactions of different orders. To attack the challenge from training the deeper and multi-layer structure of DHEN, we propose a co-designed training system that improves training efficiency of training hierarchical architectures. Evaluation over large-scale dataset from CTR prediction tasks demonstrates 0.27\% improvement on the NE of CTR prediction and 1.2x better training throughput.

\bibliographystyle{plain}
\bibliography{ref}
% \newpage
% \input{8-appendix}
\end{document}